# Multi-wavelength dissipative soliton operation of an erbium-doped fiber laser

**H. Zhang, D. Y. Tang\*, X. Wu and L. M. Zhao**

*School of Electrical and Electronic Engineering, Nanyang Technological University, Singapore, 639798*

*\*Corresponding author:* edytang@ntu.edu.sg

**Abstract:** We report on the generation of multi-wavelength dissipative soliton (DS) in an all normal dispersion fiber laser passively mode-locked with a semiconductor saturable absorber mirror (SESAM). We show that depending on the strength of the cavity birefringence, stable single-, dual- and triple-wavelength DSs can be formed in the laser. The multi-wavelength soliton operation of the laser was experimentally investigated, and the formation mechanisms of the multi-wavelength DSs are discussed.
©2009 Optical Society of America

## References and links

## 1. Introduction

Multiwavelength mode-locked fiber lasers have versatile applications, including fiber optic sensing, photonic component characterization and WDM optical communications. Several methods for achieving multiwavelength mode locking have been studied. Li et al. reported the generation of triple-wavelength picosecond mode locked pulses using a self-seeded Fabry–Perot laser diode with fiber Bragg gratings [1]. Multiwavelength actively mode-locked fiber lasers incorporating either a single sampled fiber Bragg grating [2] or a biased semiconductor optical amplifier [3] in cavity were shown by Yao et al. By virtue of the nonlinear polarization rotation (NPR) effect, simultaneous dual- and five-wavelength actively mode-locked erbium-doped fiber lasers at 10 GHz were demonstrated by Pan et al [4]. Although multiwavelength actively mode-locked fiber lasers have the advantages such as high repetition rates, narrow linewidth, they also have the drawbacks of broad pulse width, low peak power, and expensive as a modulator is required to be inserted in the cavity. Moreover, as actively mode locked multiwavelength pulses have only weak nonlinearity, they are impossible to be shaped into optical solitons that possess the born preponderance: good stability, low time jittering, short pulse width and high peak power.

Back in 1992, Matsas et al. reported the experimental observation of dual-wavelength soliton emission in a fiber laser exploiting the nonlinear polarization rotation (NPR) technique for mode locking [5]. A long laser cavity made of anomalous dispersion fibers was adopted in their experiment. However, no explanation on the formation mechanism of dual-wavelength solitons was given. In this paper we report on the experimental observation of multiple wavelength dissipative soliton operation of an erbium-doped fiber laser. Formation of dissipative solitons in normal dispersion fiber lasers has recently attracted considerable attention [6-14]. Although in an all-normal dispersion fiber laser there is no natural balance between the actions of the fiber dispersion and fiber nonlinear optical Kerr effect, therefore, no natural nonlinear Schrödinger equation soliton is formed. It was shown that as a result of the mutual interactions among the normal cavity dispersion, cavity fiber nonlinear Kerr effect, and the effective laser gain bandwidth, an optical soliton can still be formed in the laser. The formed solitons were known as the dissipative solitons [14]. Occasionally they were also called as the gain-guided solitons to distinguish from those solitons formed in the anomalous dispersion fiber lasers [5]. Single wavelength dissipative solitons have been observed in various fiber lasers [6-12]. In a previous paper we have also reported the observation of dissipative vector solitons in a dispersion managed cavity fiber laser [6]. However, to the best of our knowledge, no multiwavelength dissipative solitons operation of a fiber laser has so far been reported.

## 2. Experimental setup and experimental results

Our fiber laser is schematically shown in Fig. 1a. It has a ring cavity made of pure normal dispersion fibers. A piece of 5.0 m, 2880 ppm Erbium-doped fiber (EDF) with group velocity dispersion (GVD) of -32 (ps/nm)/km was used as the gain medium, and all the other fibers are the dispersion compensation fiber (DCF) with GVD of -4 (ps/nm)/km. The cavity has a length of 13.5m. Mode-locking of the laser is achieved with a semiconductor saturable absorber mirror (SESAM). A polarization independent circulator was used to force the unidirectional operation of the ring and simultaneously incorporate the SESAM in the cavity. Note that within one cavity round-trip the pulse propagates twice in the DCF between the circulator and the SESAM. A 50% fiber coupler was used to output the signal, and the laser was pumped by a high power Fiber Raman Laser source (KPS-BT2-RFL-1480-60-FA) of wavelength 1480 nm. The maximum pump power can be as high as 5W. All the passive components used (WDM, Coupler, and Circulator) were made of the DCF. An optical spectrum analyzer (Ando AQ-6315B) and a 350MHZ oscilloscope (Agilen 54641A) together with a 2GHZ photo-detector were used to simultaneously monitor the spectra and the mode locked pulse train, respectively. The SESAM used was made based on GaInNAs quantum wells. It has a saturable absorption modulation depth of 30%, a saturation fluence of 90 μJ/cm$^2$ and a recovery time of 10 ps. The SESAM was pigtailed with 0.5m DCF.

Mode locking of the laser self-started as the pump power was increased above the mode locking threshold. Immediately after the mode locking, multiple soliton pulses were generally formed in the cavity. However, through carefully decreasing the pump power, the number of soliton pulses could be reduced and eventually a single soliton operation state could be obtained. Fig. 2 shows a typical single soliton operation state of the laser. The optical spectrum of the pulse has the characteristic steep spectral edges, which shows that it is a dissipative soliton [6]. Based on the measured autocorrelation trace, the soliton pulse width is estimated 28.8 ps if a sech$^2$-shape pulse is assumed. Experimentally we confirmed that the soliton pulse was linearly polarized. The center wavelength of the dissipative soliton shown in Fig. 2 is located at 1576.2 nm. Experimentally, adjusting the orientations of the paddles of the PC, which corresponds to changing the linear birefringence of the cavity, the central wavelength of the soliton could be varied in a wide range from 1570nm to 1590nm. Nevertheless, solitons with central wavelengths close to 1570 or 1590nm were less stable.

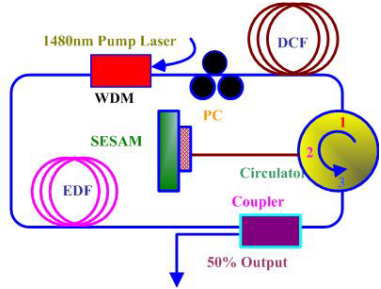

Fig. 1: Schematic of the experimental setup. EDF: Erbium doped fiber. WDM: wavelength division multiplexer. DCF: dispersion compensation fiber. PC: polarization controllers.

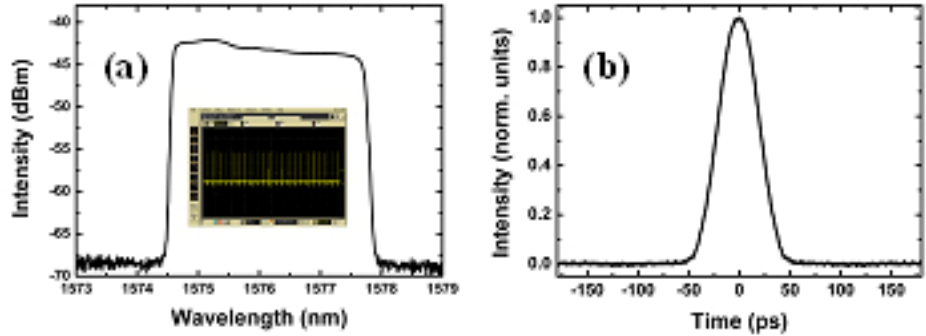

Fig. 2: (a) Optical spectra of single wavelength GGS. Insert: the oscilloscope trace. (b) The corresponding autocorrelation trace

Under stronger pumping, a new soliton was formed in the cavity. It was found that the features of the new soliton sensitively depended on the cavity birefringence. Fig. 3 shows a soliton operation state experimentally observed. Fig. 3a is the optical spectrum of the laser emission. Compared with the spectrum shown in Fig. 2, another steep-edge shaped spectrum with different central wavelength also appeared. Fig. 3b shows the oscilloscope trace of the laser emission. There are two solitons in the cavity. Each soliton has different pulse energies as represented by the two different pulse heights in the oscilloscope trace. The two solitons also propagated with different group velocities in the cavity. Therefore, when triggered with one soliton the other soliton then moved randomly on the oscilloscope screen, indicating that the two solitons have different group velocities in the cavity. We had further experimentally studied the polarization features of the pulses using an external cavity polarizer. Through rotating the external cavity polarizer it was identified that one soliton could be completely suppressed while the other still remained on the oscilloscope trace, associated with the suppression of one soliton pulse on the oscilloscope one squared-shaped spectrum also disappeared on the optical spectrum. Fig. 3 actually shows a state of the dual wavelength dissipative soliton operation of the fiber laser. Specifically, the dissipative soliton with the large pulse energy has a center wavelength of 1576.2nm, and the one with the weak energy has a center wavelength of 1579.6nm, and the two dissipative solitons have orthogonal polarizations.

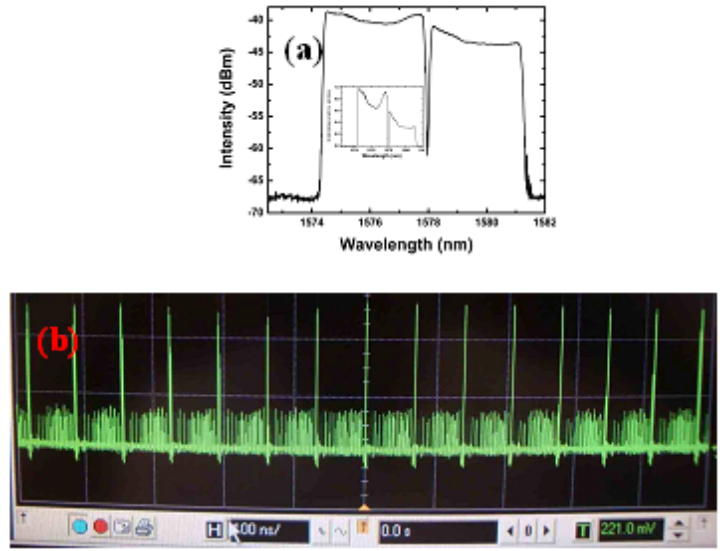

Fig. 3: (a) Optical spectrum of dual wavelength GGSs. Insert: the normalized optical spectrum;
(b) Oscilloscope trace of dual wavelength GGSs.

The relative strength of the solitons varied with the cavity birefringence. Slightly tuning the orientation of the PC, one could continuously change the relative soliton pulse energy. At a certain soliton intensity relation, it was found that two solitons could even have the same group velocity despite of the fact that they have different central wavelengths. Fig. 4 shows the oscilloscope of such a case, where the two solitons have fixed soliton separation as they circulated in the cavity. However, such a state was unstable, after a short time the solitons lost their synchronization and moved independently again.

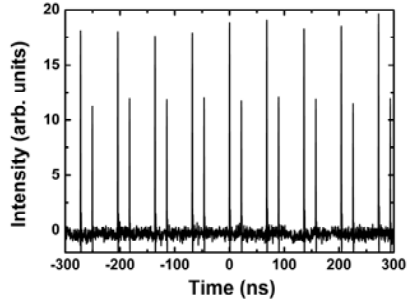

Fig. 4: Oscilloscope trace of synchronized dual wavelength GGS.

Through Careful control of the cavity birefringence, experimentally we found that the dual wavelength solitons could also be transformed into a single wavelength vector dissipative soliton, as shown in Fig. 5. To obtain the result we have significantly changed the orientation of the PC paddles while the pump strength was kept fixed. On the oscilloscope trace the previous two solitons now merged together, consequently only one pulse could be observed in the cavity. Checked with a high speed oscilloscope (50GHz) combined with a commercial autocorrelator (FR-103MN), no fine structures were detected within the pulse, confirming it is vector soliton. Note that the central wavelength of the vector dissipative soliton has now shifted to a value between those of the linearly polarized solitons shown in Fig. 3a. The polarization resolved study on the state further confirmed that the soliton pulse was elliptically polarized, and its two orthogonal components have comparable spectral intensity and the same central wavelength [7].

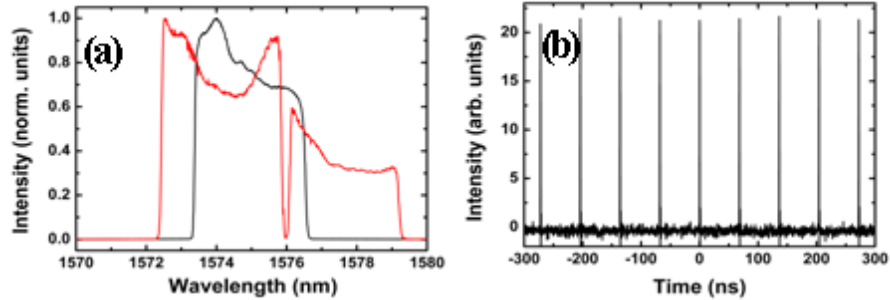

Fig. 5: (a) Optical spectra of polarization locked gain guided vector soliton and dual wavelength spectrum obtained through rotating PCs but kept the pump strength fixed: normalized unit. (b) Oscilloscope trace of polarization locked gain guided vector soliton after passing through a polarizer.

Besides the above dual-wavelength dissipative soliton operation, experimentally under certain laser operation conditions, triple-wavelength dissipative solitons as shown in Fig.6 were also observed. Experimental studies turned out that all the solitons were linearly polarized, and the two solitons with the shortest and longest wavelength, respectively, have the same polarization, which is orthogonal to the soliton that has the middle value of the central wavelength.

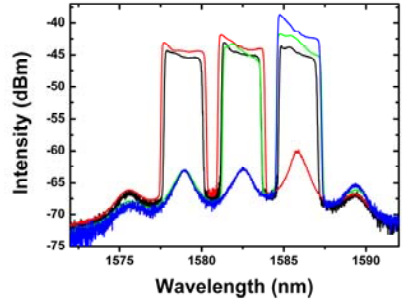

Fig. 6: Single/dual/triple wavelength spectra obtained through rotating PCs but kept the pump strength fixed.

## 3. Formation mechanisms of the multi-wavelength dissipative solitons

We have experimentally investigated the formation mechanism of the multiple wavelength solitons in our fiber laser. Starting from a state of dual-wavelength soliton operation as shown in Fig. 2, slowly reducing the pump strength, it was found that only the spectral intensity became weaker, while the changes on other soliton parameters, such as the spectral bandwidth, central wavelength and pulse duration were not obvious. As the pump power became about 110 mw, one of the soliton disappeared and the laser emission jumped to a single wavelength soliton operation state. If the pump power is further reduced till the mode locking is destroyed, continuous laser emission either at the 1576 nm or the 1580 nm, or simultaneously at both wavelengths could be observed. The CW laser emissions have the same polarization as those of the corresponding solitons. This experimental result shows that the dual-wavelength soliton emission of the laser should be related to the birefringence of the cavity. Similar feature has also been observed on the triple–wavelength soliton emission as have been shown in Fig. 5. Decreasing pump power until mode locking was destroyed, on the spontaneous emission spectrum of the laser one could clearly see the existence of a periodic spectral intensity modulation, indicating the existence of a spectral filter in the cavity, which could also be traced back to the cavity birefringence effect.

Based on these experimental evidences and features of the multi-wavelength solitons, we explain the formation mechanism of the dual- and triple-wavelength solitons as the following: despite of the fact that no polarizing components were used in the cavity, the slight residual polarization asymmetry of the components used, such as the SESAM and the circulator, could still cause the formation of a linear artificial birefringent filter in the cavity. It is known that the wavelength spacing between the transmission maxima of a birefringence filter depends on the cavity birefringence, which is given by $\Delta\lambda= \lambda^2/(LB)$, where $\lambda$ is the central wavelength, L is the cavity length and B is the strength of birefringence [16]. The stronger the cavity birefringence, the smaller is the spacing. The existence of the artificial birefringence filter results in the multiple wavelength lasing of the laser, and the saturable absorption effect of the SESAM further causes the self-started mode locking of the laser In the case of strong cavity birefringence, the artificial filter has narrow bandwidth, within the effective laser gain bandwidth range, three wavelength lasing and mode locking is possible. . However, due to the polarization gain competition among the solitons, solitons with wavelength separation smaller than the homogeneous gain bandwidth cannot have the same polarization but orthogonal polarizations, while solitons with wavelength separation larger than the homogeneous gain bandwidth could simultaneously exist. With smaller cavity birefringence the filter bandwidth also becomes broader. Within the effective gain bandwidth, eventually only solitons with two different central wavelengths could be excited. As a result of the polarization gain competition the two soliton could not have the same polarization. Therefore, dual-wavelength soliton with orthogonal polarizations were observed. When the birefringence of the cavity is further decreased, the bandwidth of the filter became so broad that only a single wavelength soliton could be formed in the laser, in this case the cross-coupling between the two polarization-components of the cavity also became strong. As a result of the strong cross-phase modulation vector dissipative soliton could be formed in the laser. Based on this explanation we found the results of our experimental observations could be well understood.

## 4. Conclusion

In conclusion, we have experimentally observed multiwavelength dissipative soliton operation of an erbium-doped fiber laser mode locked with a SESAM. It was shown that depending on the strength of the linear cavity birefringence either dual- or triple-wavelength soliton operations could be obtained. We found that the multi-wavelength soliton operation of the laser could be well explained as caused by the existence of an artificial birefringence filter in the laser cavity. The multiple wavelength passively mode locked fiber laser may find applications in the fiber sensors or optical signal processing systems.

**Acknowledgement**

This project is supported by the National Research Foundation Singapore under the contract NRF-G-CRP 2007-01.